\documentstyle[12pt,aas2pp4]{article}
\def\insfig#1{#1}
\def\endinsfig{\end{document}}

\font\smallrm=cmr8

\def\FWHM{{\smallrm FWHM}}

\def\kms{\hbox{$\,$km$\,$s$^{-1}$}}

\def\gtorder{\mathrel{\raise.3ex\hbox{$>$}\mkern-14mu
             \lower0.6ex\hbox{$\sim$}}}
\def\ltorder{\mathrel{\raise.3ex\hbox{$<$}\mkern-14mu
             \lower0.6ex\hbox{$\sim$}}}

\lefthead{Tonry}
\righthead{Gravitational Lens Redshifts}

\begin{document}

\title{Redshifts of the Gravitational Lenses MG~1131+0456 and B~1938+666\altaffilmark{1}}

\author{John L. Tonry}
\affil{Institute for Astronomy, University of Hawaii, Honolulu, HI 96822}
\affil{Electronic mail: jt@avidya.ifa.hawaii.edu}
\authoremail{jt@avidya.ifa.hawaii.edu}

\author{Christopher S. Kochanek}
\affil{Harvard-Smithsonian Center for Astrophysics, Cambridge, MA 02138}
\affil{Electronic mail: kochanek@cfa.harvard.edu}
\authoremail{jt@avidya.ifa.hawaii.edu}

\altaffiltext{1}{Based on observations at the W. M. Keck Observatory,
which is operated jointly by the California Institute of Technology
and the University of California}

\begin{abstract}
The redshifts of the gravitational lens galaxies in MG~1131+0456 and
B~1938+666 are $0.844$ and $0.881$ respectively.  Both are early-type
galaxies lying at the redshifts predicted by assuming that they are
early-type galaxies with old stellar populations lying on the 
fundamental plane.  We also find evidence for a foreground group
of galaxies at $z=0.343$ near MG~1131+0456.  The source
redshifts are predicted to be $\gtorder 1.8$ in both systems, but they are so
red that infrared spectra will be required to determine their redshifts.

\noindent \it{Subject headings:} cosmology --- distance scale ---
gravitational lensing 
\end{abstract}

\section{Introduction}

There are now over 50 multiple-image gravitational lens systems.  The
lens sample is a unique and powerful tool for studying cosmology,
galactic structure, and galactic evolution.  For studying galaxy
structure and evolution: (1) it is the only sample of galaxies
selected based on mass rather than luminosity; (2) it is one of the
largest samples of galaxies with precisely measured masses; (3) it is
comparable in size to the largest kinematic samples outside the local
universe.  The lenses are one of the more promising alternatives
to measuring
$H_0$ other than using the local distance ladder (see Koopmans \& Fassnacht
1999), and their incidence and redshift distributions can be used to
measure the underlying geometry ($\Omega_0$, $\Lambda_0$, see Falco,
Kochanek \& Munoz 1998).  They can also be used to study the hosts of
high-redshift AGN and quasars (Rix et al. 1999) and the dust (Falco et
al. 1999a) and interstellar medium (e.g. Menten, Carilli \& Reid 1998)
of the lens galaxy.  Surveys to find new lenses are now very
productive (e.g. Browne et al. 1999), and HST imaging surveys are
providing detailed surface photometry for most of the lens galaxies
(Falco et al. 1999b).  Thus, the remaining barrier to exploiting the
astrophysical potential of gravitational lenses is the high fraction
of unmeasured lens and source redshifts.  In this paper we determine
the lens galaxy redshifts of MG~1131+0456 and B~1938+666.

MG~1131+0456 (Hewitt et al. 1988) is a radio-selected lens, and it
was the first Einstein ring to be
discovered.  The radio map (Chen \& Hewitt 1993) shows two images of the
radio core, an elliptical ring image formed from the jet, an unlensed 
radio lobe, and a central, unresolved image at the position of the lens.
Ground-based optical (Hammer et al. 1991) and infrared (Annis 1992, 
Larkin et al. 1994) established that the system was very red, but HST
imaging was required to understand the morphology (Kochanek et al. 1999a).
At optical wavelengths the emission is 
dominated by a red, early-type lens galaxy with a photometric redshift
of  $0.84 < z_l < 1.01$ (Kochanek et al. 1999b).  There is a faint,
incomplete optical ring image of the host galaxy of the radio AGN.  
In the infrared, the H-band image is dominated by a bright, complete
Einstein ring image of the AGN host galaxy (H$\simeq17.4$ mag). The
AGN host appears to be a very dusty galaxy, which would
be classified as an ERO (``extremely red object'', e.g. Thompson et al. 1999), while the
lens galaxy is a normal, transparent early-type galaxy.  Hammer et al.
(1991) proposed a redshift of $0.85$ for the lens galaxy. 

B~1938+666 (King et al. 1997) was also radio-selected.  The radio map
shows a partial Einstein ring formed from four images of the radio jet
and two offset images of the radio core.  Like MG~1131+0456, ground
based optical and infrared imaging found that the system was very 
red (Malhotra, Rhoads \& Turner  1997)
The lens galaxy again appears to be a red, 
early-type galaxy with a photometric redshift of $0.81 < z_l < 1.04$,
with no sign of any optical images of the source (Kochanek et al. 1999b).  
In the infrared, roughly equal amounts of flux come from the
lens galaxy and an almost perfectly circular Einstein ring image
of the AGN host galaxy (King et al. 1998).
      
By modern standards, neither lens galaxy is faint (I=$20.8$ and $21.5$
mag respectively).  The measurement is difficult only because the
spectral features of early-type galaxies at $z \gtorder 0.7$ are
buried in the Meinel OH bands.  The source redshifts, however, are
almost certainly unmeasurable in the optical -- the emission is
very faint and dominated by the host galaxy rather than an AGN.
In \S2 we report measurements of the lens redshift in MG~1131+0456 and
in B~1938+666.  We discuss the
consequences of the measurements in \S3.

\section{Observations and Reductions}

MG~1131+0456 was
observed on 11 and 12 May 1999 using the Low Resolution Imaging
Spectrograph (LRIS, Oke et 
al. 1995) at the Keck-2 telescope on Mauna Kea.
Both nights were clear and the seeing was variable,
but extremely good ($<0\farcs5$) on the night of 11 May.  We
used a 1\farcs0 slit combined with the 400~l/mm grating blazed
at 8500\AA, to obtain spectral coverage from 6250--10050\AA with a 
spectral resolution $\simeq 5$\AA\ \FWHM\ and a
scale along the slit of 0\farcs211/pixel.  
The MG~1131+0456 observation totaled 3900 seconds on
the lens, and serendipitously caught three other galaxies, G1, G2, and
G3, at the same time.  The slit was oriented at a position angle of 
$63^\circ$, as illustrated in Figure 1.

B~1938+666 was observed on 26 and 27 October 1997, and 
again on 23 July 1998.
The October nights had light cirrus and $\sim$1\arcsec\ seeing;
July was clear and had excellent seeing of about 0\farcs6.
In October 1997 the grating was rotated to provide coverage from
3800--8700\AA\ with a spectral resolution $\simeq 7.9$\AA\ \FWHM;
in July 1998 the 400~l/mm grating was used over the range
5250--9050\AA with a spectral resolution $\simeq 5$\AA\ \FWHM.
The much improved seeing in July resulted in much more signal for the
lensing galaxy, so the analysis of the lens galaxy is based only on those
data.  Figure 1 shows the two slit positions which were used.  At
PA~96 (October 1997) the slit caught galaxy G2 which showed [OII] emission,
and at PA~133 (July 1998) the slit also caught galaxy G1.

The observing program was similar in all respects to that of Tonry
(1998), and the same template stars were used for the redshift
determinations reported here.  A log of the observations is presented
in Table 1.

\insfig{
\begin{figure}
\epsscale{1.0}
\caption[fig1.eps]{
Illustration of the slit positions and galaxy identifications for
MG~1131+0456 and B1938+666.  North and East are indicated in each
diagram as well as a 10\arcsec\ scale.
\label{fig1}}
\end{figure}
}

The spectra were reduced using software described in detail by Tonry
(1984).  The basic steps are to flatten the images, remove cosmic
rays, derive a wavelength solution as a function of both row and
column using sky lines (wavelengths tabulated by Osterbrock et
al. 1996), derive a slit position solution as a function of both row
and column using the positions of the template star images in the
slit, rebin the entire image to coordinates of log wavelength and slit
position, add images, and then sky subtract.  A linear fit to
patches of sky on either side of the object (including a patch between
for the galaxy pair observations) did a very good job of removing the
sky lines from the spectra.
In each case the spectrum was extracted, and
cross-correlated with the template spectrum according to Tonry and
Davis (1979), as well as being analyzed by the Fourier quotient method
of Sargent et al. (1977).  The cross-correlation is more robust in
the case of low signal to noise, but at the signal levels here the two
results are statistically the same, and are simply averaged.  For
each spectrum the redshift and error were calculated.
Table 2 lists the redshifts, errors,
and cross-correlation significance ``$r$'' values for each spectrum or
emission lines used for redshift determination.

\subsection{MG~1131+0456}

The spectrum of the lens galaxy in MG~1131 is that of a 
typical early-type galaxy at a 
redshift of $z_l=0.8440\pm0.0005$.  With a cross correlation $r$ value of
7.6, the redshift is determined to a high degree of confidence, and it is
confirmed by the H and K lines visible at 7300\AA\ and the weak [OII]
emission.  The spectrum is illustrated in Figure 2.  No hint of source
galaxy features could be discerned in the spectrum.

\insfig{
\begin{figure}
\epsscale{1.0}
\plotone{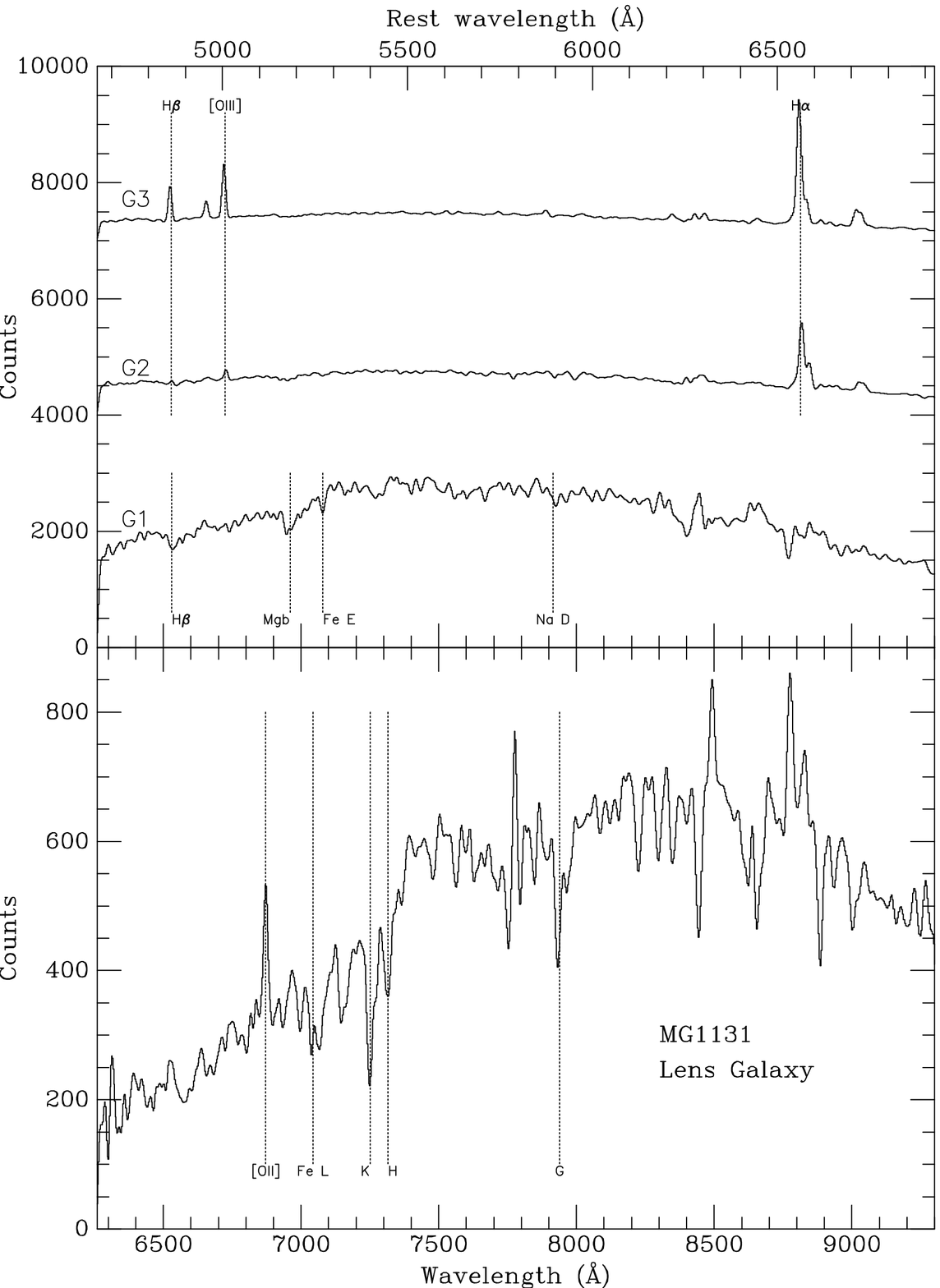}
\caption[fig2.eps]{ The spectra of the lensing galaxy in
MG~1131+0456 and the neighboring galaxies G1, G2, and G3 are shown.
The spectra were smoothed with a \FWHM\ 9\AA\ Gaussian.
The top axis shows rest wavelength at the redshift of 0.343 of
the neighboring groups, and some emission and absorption lines are
marked.  The lensing galaxy is shown in the bottom panel, and various
absorption lines and the [OII] 3727 line are marked.
Most of the prominent features beyond 8400\AA\ are
poorly subtracted sky lines.
The atmospheric A and B bands have been divided out of the spectrum.
For clarity, the spectra of G2 and G3 have been divided by 3 and offset by 4000 and
7000 counts respectively.
\label{fig2}}
\end{figure}
}

The other three galaxies caught in the slit, G1, G2, and G3, are part
of a group or cluster at redshift 0.343; G1 is an early-type galaxy, whereas
the spectra of G2 and G3 are dominated by very strong 
emission lines.\footnote{In the labeling scheme used by Kochanek et al. (1999a),
our galaxy G1 is their galaxy G2, and our galaxies G2 and G3 lie outside
the field of view of their HST images.}  The
velocity dispersion of the cluster inferred from these three galaxies
is only 233\kms, but it seems likely that this is quite a rich group
given that the slit happened to pick up three at once.  This is in 
addition to the group or cluster which seems to be associated with the
lens galaxy (see Kochanek et al. 1999a)

\subsection{B~19381+666}


The poor seeing in the 1997 observing run precluded determination of a
redshift for the B1938 lensing galaxy, although features which might
be construed to be H and K were found at $z = 0.88$.  The galaxy G2,
which was placed in the slit at PA~96, has a single, strong emission
line which is almost certainly [OII] at a redshift of 0.8784, judging
from its width and isolation.  Revisiting the lensing galaxy in 1998
under conditions of good seeing soon provided a spectrum which shows
that it has a redshift of $z_l=0.881\pm0.0005$.  These spectra are
plotted in Figure 3.  Again, no hint of source galaxy features could
be spotted.

We list in Table 2 a tentative redshift of $0.924$ for galaxy G1, which lay in the
PA~133 slit, but the $r$ value of only 3.9 makes it very possible that
this redshift is not correct.

\insfig{
\begin{figure}
\epsscale{1.0}
\plotone{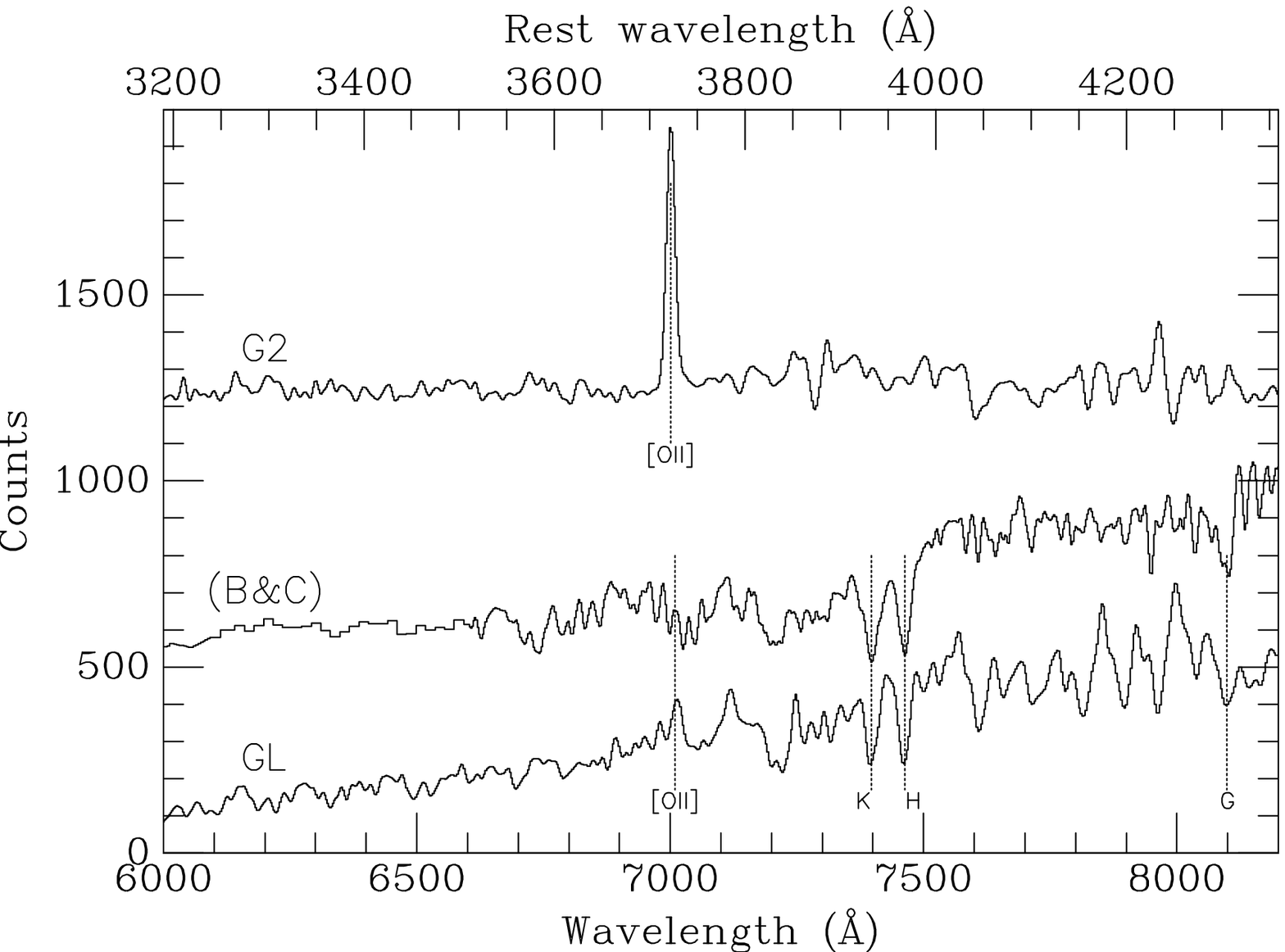}
\caption[fig3.eps]{The spectra are shown for G2 and the lensing
galaxy in the B~1938 system.
These spectra have been Gaussian smoothed with a \FWHM\ of 9\AA.  The
top axis shows rest wavelength at a redshift of 0.881.  A single
strong emission line in G2 is probably [OII] 3727, and it is also
faintly visible in GL.  The middle curve is a Bruzual \& Charlot 1997
spectrum of a 10~Gyr single burst model, and illustrates how many of
the UV features seen in GL between O[II] and Ca K are real.
For clarity, G2 has been offset by 1100 counts.
\label{fig3}}
\end{figure}
}

\section{Discussion}

The MG~1131+0456 and B~1938+666 lens galaxies are normal, passively evolving
early-type galaxies at $z_l=0.844$ and $0.881$ respectively. We see this
directly in their spectra.  The spectroscopic redshifts agree very well
with the photometric redshift estimates of Kochanek et al. (1999b) based
on the fundamental plane of lens galaxies and the agreement of the broad
band colors with passively evolving models of stellar populations.

With the lens redshift known, we can use the fundamental plane to estimate
the source redshift.  The limits will be imprecise because the position of
the lens galaxy relative to the fundamental plane depends on the source redshift 
only through the effects of the weakly varying distance ratio $D_{LS}/D_{OS}$ 
on the inferred velocity dispersion of the lens galaxy (see Kochanek et al. 1999b).  
The estimated velocity dispersion, $\sigma^2 \propto D_{OS}/D_{LS}$, is divergent
at the lens redshift and monotonically declines with increasing source redshift.
Thus, we obtain firm lower bounds of  $z_s>1.9$ and $z_s>1.7$ for 
the source redshifts of MG~1131+0456 and B~1938+666 respectively.  There is no
firm upper bound for the redshift of either source over the range we considered 
($1<z_s <5$).  For MG~1131+0456 there is no statistically significant estimate
for a best source redshift, while for B~19388 we estimate that $z_s=2.8$.
In both systems, the optical counterparts of the source are too faint for redshift
measurements, but both should be measurable with the new generation
of infrared spectrographs.  MG~1131+0456 is very bright in the infrared
and the counterparts of the radio cores become steadily brighter 
at longer wavelengths (see Larkin et al. 1994) as the dust in the host 
galaxy becomes increasingly
transparent.   B~1938+666 is more challenging because it is fainter and lacks 
obvious counterparts to the radio cores.  Both hosts are examples of extremely
red galaxies which probably would be missed in surveys based on the ``UV-dropout''
method.

\acknowledgements
As always, thanks are due to the scientists and engineers responsible 
for the Keck telescope and the LRIS spectrograph.  We also thank E.
Falco for reading the manuscript.

\clearpage

\begin{deluxetable}{rlrrrrrrr}
\tablecaption{Observing Log.\label{tbl1}}
\tablewidth{0pt}
\tablehead{
\colhead{Obs\#} &  \colhead{Objects} & \colhead{UT Date}  &
\colhead{UT}  & \colhead{$\sec\,z$} & 
\colhead{PA} & \colhead{Exposure} & 
\colhead{Grating}
} 
\startdata
 1. & B~1938   & 26/10/97 & 04:59 & 1.18 &  96 & 1500 & 300/5000 \nl
 2. & B~1938   & 26/10/97 & 05:25 & 1.44 &  96 & 1500 & 300/5000 \nl
 3. & B~1938   & 27/10/97 & 04:47 & 1.08 & 133 & 1500 & 300/5000 \nl
 4. & B~1938   & 27/10/97 & 05:15 & 1.38 & 133 & 1500 & 300/5000 \nl
 5. & B~1938   & 23/07/98 & 06:04 & 1.97 & 133 & 1500 & 400/8500 \nl
 6. & B~1938   & 23/07/98 & 06:33 & 1.83 & 133 & 1500 & 400/8500 \nl
 7. & B~1938   & 23/07/98 & 07:01 & 1.73 & 133 & 1500 & 400/8500 \nl
 8. & B~1938   & 23/07/98 & 07:31 & 1.63 & 133 & 1500 & 400/8500 \nl
 9. & MG~1131  & 11/05/99 & 06:50 & 1.03 &  63 & 1200 & 400/8500 \nl
10. & MG~1131  & 11/05/99 & 07:11 & 1.04 &  63 & 1200 & 400/8500 \nl
11. & MG~1131  & 11/05/99 & 06:29 & 1.03 &  63 & 1500 & 400/8500 \nl
\enddata
\tablecomments{PA is east from north, exposures are in seconds.}
\end{deluxetable}

\begin{deluxetable}{lrrrc}
\tablecaption{Redshifts and Emission Lines.\label{tbl2}}
\tablewidth{0pt}
\tablehead{
\colhead{Galaxy} & \colhead{$y$} & \colhead{$z$} & \colhead{$\pm$}  &
\colhead{$r$/emission}
} 
\startdata
MG~1131 GL & $0.0$  & 0.8440 & 0.0005 & 7.6 {[OII]} \nl
MG~1131 G1 &$15.6$  & 0.3438 & 0.0002 & 5.0 \nl
MG~1131 G2 &$31.0$  & 0.3437 & 0.0002 & {[OII]$^*$,H$\beta$,[OIII],H$\alpha$,[NII],[SII]} \nl
MG~1131 G3 &$75.3$  & 0.3420 & 0.0002 & {[OII]$^*$,H$\beta$,[OIII],H$\alpha$,[NII],[SII]} \nl
B~1938  GL & $0.0$  & 0.8809  & 0.0005  & 5.9 {[OII]?}\nl
B~1938  G1 & $4.4$  & 0.924?  & 0.001   & 3.9\nl
B~1938  G2 & $9.5$  & 0.8784  & 0.0005  & [OII]\nl
\enddata
\tablecomments{Columns: 
Galaxy name, slit position (\arcsec), redshift and error,
cross-correlation $r$ value or emission lines ($^*$ indicates second order)}
\end{deluxetable}

\endinsfig

\clearpage

\centerline{\bf FIGURE CAPTIONS}
\bigskip

\figcaption[fig1.eps]{
Illustration of the slit positions and galaxy identifications for
MG~1131+0456 and B~1938+666.  North is up and East is left in each
picture and a 10\arcsec\ scale is given.
\label{fig1}}

\figcaption[fig2.eps]{ The spectra of the lensing galaxy in
MG~1131+0456 and the neighboring galaxies G1, G2, and G3 are shown.
These spectra has been Gaussian smoothed with a \FWHM\ of 
9\AA.  The top axis shows rest wavelength at the redshift of 0.343 of
the neighboring groups, and some emission and absorption lines are
marked.  The lensing galaxy is shown in the bottom panel, and various
absorption lines and the [OII] 3727 line are marked.
Most of the prominent features beyond 8400\AA\ are
poorly subtracted sky lines.
The atmospheric A and B bands have been divided out of the spectrum.
For clarity, G2 and G3 have been divided by 3 and offset by 4000 and
7000 counts respectively.
\label{fig2}}

\figcaption[fig3.eps]{ The spectra are shown for G2 and the lensing
galaxy in the B~1938 system.
These spectra have been Gaussian smoothed with a \FWHM\ of 9\AA.  The
top axis shows rest wavelength at a redshift of 0.881.  A single
strong emission line in G2 is probably [OII] 3727, and it is also
faintly visible in GL.  The middle curve is a Bruzual \& Charlot 1997
spectrum of a 10~Gyr single burst model, and illustrates how many of
the UV features seen in GL between O[II] and Ca K are real.
For clarity, G2 has been offset by 1100 counts.
\label{fig3}}

\end{document}